\documentclass[epj,final]{svjour} 

\usepackage{graphics}
\usepackage{epsf}
\usepackage{epsfig}

\begin{document}

\flushbottom
\def\bottomfraction{0.5}


\title{A deterministic sandpile automaton revisited}

\author{S.~L\"ubeck\inst{1}      \thanks{E-mail:~sven@thp.uni-duisburg.de} 
        \and N.~Rajewsky\inst{2} \thanks{E-mail:~rajewsky@math.rutgers.edu}
        \and D.~E.~Wolf\inst{1}  \thanks{E-mail:~wolf@rs1.comphys.uni-duisburg.de}
       }

\institute{Theoretische Physik, Gerhard-Mercator-Universit\"at Duisburg, 
  47048 Duisburg, Germany 
  \and
  Department of Mathematics, Rutgers University,
  New Brunswick, New Jersey 08903, USA}

\date{Received: 16 August 1999}

\abstract{
The Bak-Tang-Wiesenfeld (BTW) sandpile 
model is a cellular automaton which has been intensively 
studied during the last years as a
paradigm for self-organized criticality. 
In this paper, we reconsider a deterministic version of the BTW 
model introduced by Wiesenfeld, Theiler and McNamara, 
where sand grains are added always to one
fixed site on the square lattice. 
Using the {\it Abelian sandpile} formalism we discuss 
the static properties of the system.
We present numerical evidence that the deterministic
model is only in the BTW universality class
if the initial conditions and the geometric
form of the boundaries do not respect the full
symmetry of the square lattice.
}


\maketitle

\section{Introduction}

In equilibrium systems with short-ranged interactions
and no broken continuous symmetry,
correlation functions
usually decay exponentially with the distance.
Exceptions from this behavior occur only in special
cases, for example at critical points in a phase diagram.
In this case the corresponding correlation
functions decay algebraically if 
the relevant system parameter is adjusted
to special (critical) values~(e.g.~critical temperature, critical pressure, 
etc.).

The concept of self-organized criticality (SOC)
which was introduced by Bak, Tang and Wiesenfeld 
in 1987~\cite{BAK_1,BAK_2} 
attempts to explain the fact that in 
nature critical behavior
is often observed, although nature cannot ''fine-tune
parameters'' (see~\cite{GRINSTEIN_1,BAK_3} for an 
introduction and overview). The term ``critical behavior''
corresponds here to a power-law behavior of the
probability distributions of certain physical quantities 
which characterize the system in both
space and time~\cite{BAK_1,BAK_2}. Typical examples for such
quantities are the size and life times of catastrophic events.
The main idea is then that the critical state is an attractor
of the dynamics.

One of the paradigmatic systems which exhibit SOC is the
Bak-Tang-Wiesenfeld sandpile model (BTW model)
which was intensively investigated in the
past. 
In the following, we restrict our discussion to the BTW model on
the two-dimensional square lattice. 
Concerning static properties, analytical results
exist which are mainly due to Dhar, who 
developed a formalism for {\it Abelian 
sandpile models}~\cite{DHAR_2}
which allows to calculate exactly
the height probabilities, height 
correlations, number of steady state configurations, 
etc.~\cite{DHAR_2,MAJUM_1,PRIEZ_1,IVASH_1}.
However, much less is known rigorously about the dynamical features,
and estimates for the exponents of the probability distributions 
of avalanche quantities are only known from computer simulations~(see
for instance \cite{GRASS_1,MANNA_1,LUEB_2,LUEB_4,CHESSA_1,LUEB_5});
further guesses for these exponents have been made
via renormalization group approaches (e.g.~\cite{OBUKHOV_1,DIAZ_2}).

In this paper we consider a deterministic version of
the BTW model which was introduced
by Wiesenfeld {\it et al.}~\cite{WIESENFELD_1}. 
In the original BTW model, sand grains are 
added at randomly chosen sites of the lattice. 
In the deterministic BTW model (DBTW) the seeding of sand
is confined to one special site of the lattice.
Computer simulations revealed \cite{WIESENFELD_1} 
that the DBTW model still displays
criticality. Therefore, randomness in the location of the 
perturbations is 
not a necessary ingredient for SOC~\cite{WIESENFELD_1}.
Furthermore, the authors concluded from their numerical analysis 
that the different versions of the BTW model could 
display different scaling behavior.

However, this conclusion was obtained from an
investigation of the DBTW model for a 
small system size.
As finite-size effects have been shown to affect
the scaling behavior of the BTW model
strongly~\cite{MANNA_1,LUEB_2}, we reinvestigate
the case here and present a systematic finite-size analysis.
We also present some new exact results for 
the DBTW model.

The paper is organized as follows. 
In Sec.~\ref{sec:definition}, we define the model and, 
using essentially the formalism developed by Dhar~\cite{DHAR_2}, 
study the static properties of the DBTW. 
In Sec.~\ref{sec:dynamics}, we present our results 
from computer simulations and
then discuss the observed scaling behavior in
the context of the universality hypothesis of Ben-Hur
and Biham~\cite{BENHUR}.
A summary closes the paper.

\section{Definition of the model and static properties}
\label{sec:definition}

The BTW model on a two-dimensional square lattice 
of size~$L\times L$ is defined as follows. To 
each site $i$ ($i=1,2,\ldots,$ $L^2$) 
an integer variable $z_i$ (the height) is assigned. 
Starting with an empty lattice ($z_i=0$ for all $i$) 
the addition of a grain of sand means
to choose a site $i$ at random and to
increase $z_i$ by $1$. 
If $z_i$ is equal to a fixed threshold value
$z_{c}$, site $i$ topples and distributes 
one grain of sand to each nearest
neighbor, which can in turn trigger more toppling events. 
Thus, an avalanche of relaxation events may take place.
For the sake of simplicity (and without loss of generality) 
we set $z_c=4$ throughout this work.
The toppling rules can be formulated in terms of a $N \times N$ toppling 
matrix~$\Delta$~\cite{DHAR_2}, where $N=L^2$.
If site $i$ topples, one has
\begin{equation} 
z_j \longrightarrow z_j - \Delta_{ij}
\label{eq:relax_rule_matrix}
\end{equation}
for all $j=1,2,...,N$, where the toppling matrix
satisfies the conditions $\Delta_{ii} = 4$,
$\Delta_{ij} = -1$ if $i$ and $j$ are nearest
neighbors, and $ \Delta_{ij} = 0$ otherwise. 
Sand falling over the rim of the system is discarded.

In the following we briefly describe the {\it Abelian
sandpile} formalism which was introduced by Dhar~\cite{DHAR_2}
and recall the major results.
Dhar showed that the dynamics of the BTW model is well defined in
the sense that the resulting 
stable configuration $C=\{z_i\}$ is always the same,
regardless of the order, in which critical sites are updated 
during an avalanche. 
One can define an operator $a_i$ by its action onto a stable
configuration $C$: $a_iC$ is the stable configuration
which results from  adding a particle at site $i$ and relaxing the
resulting configuration. Dhar showed that
\begin{equation} 
[a_i,a_j] = 0, 
\end{equation}
for all $i,j$. 
Therefore, the BTW model is called an {\it Abelian sandpile 
model}~\cite{DHAR_2}.

One is interested in the stationary state of this
cellular automaton, i.e., one iterates the dynamical rules until
all expectation values become time independent. 
Since the dynamics can be described as a Markovian process
a stable configuration can either be transient or recurrent. 
A recurrent configuration $C$ can be defined by demanding that 
for every possible seeding site~$i$
a natural number $m_i(C)$ exists such that $a_i^{m_i(C)}C = C$ holds.
Thus, for a recurrent configuration  $C$ and for a natural number $l$ 
one gets~\cite{DHAR_2}
\begin{equation} 
a_i^lC = a_i^l a_i^{m_i(C)}C = a_i^{m_i(C)}a_i^l C, 
\end{equation}
which shows that $a_i^lC$ is a recurrent configuration, too. 
Therefore, the set of recurrent configurations is closed
under the action of the operators $a_i$~\cite{DHAR_2}.
Since it is possible to define unique, inverse operators
$a_i^{-1}$, it follows that each recurrent configuration
has the same probability to appear in the stationary state.
One can further prove that the number of recurrent 
configurations is equal to~$\det{\Delta}$~\cite{DHAR_2}.

A two point correlation function $G_{ij}$ can be 
defined in the following way: 
let $G_{ij}$ be the expectation value for the number of
topplings in $j$, which are caused by adding a particle at site $i$.
In the stationary state, the average number of particles which
enter site $j$ must be equal to the average of sand grains leaving
site $j$:
\begin{equation}\label{T17}
G_{ij}\Delta_{jj}=\sum \limits_{k \not =j } G_{ik} (-\Delta_{kj})
+{\delta}_{ij}\;, 
\end{equation}
which implies $G={\Delta}^{-1}$ \cite{DHAR_2}. 
The analytic expression for
$G$ is well known:
\begin{eqnarray}
& & G_{i j} = {\Delta^{-1}}_{i j}  = \nonumber \\
& &\frac{1}{{(L+1)}^{2}} 
 \sum \limits^{L}_{a,b=1}
\frac{
\sin{x_i \tilde{a}}\, 
\sin{x_j \tilde{a}}\, 
\sin{y_i \tilde{b}}\, 
\sin{y_j \tilde{b}}\, 
}
{\sin^2{\frac{\tilde{a}}{2}} + \sin^2{\frac{\tilde{b}}{2}}},
\label{eq:corr}
\end{eqnarray}
with $\tilde{a}=a \pi /(L+1)$, $\tilde{b}=b \pi /(L+1)$ and
where the sites $i$ and $j$ have the coordinates
$(x_i,y_i)$ and $(x_j,y_j)$, respectively.
Let $s$ denote the size of an avalanche, i.e., 
the total number of topplings during that avalanche. 
Using Eq.~(\ref{eq:corr}) it can be shown that the average
number of topplings~$\langle s \rangle$ is given by 
\begin{eqnarray}
\langle s \rangle & = & \frac{1}{L^{2}}\sum \limits_{i,j} G_{i j} \nonumber \\
& = & \frac{1}{L^{2}{(L+1)}^{2}}\sum\limits^{L}_{{{a,b} \atop \mbox{\tiny odd}}}
\frac{\cot^2{\frac{\tilde{a}}{2}} \; \cot^2{\frac{\tilde{b}}{2}}}
{\sin^2{\frac{\tilde{a}}{2}} + \sin^2{\frac{\tilde{b}}{2}}}
\label{eq:meansize_1}
\end{eqnarray}
which scales for large~$L$ as~\cite{DHAR_2}
\begin{equation}
\langle s \rangle \sim L^2.
\label{eq:meansize_2}
\end{equation}

Let us now turn to the deterministic BTW (DBTW)
model as introduced by Wiesenfeld, Theiler, and 
McNamara~\cite{WIESENFELD_1}. 
Here, sand is always added at a fixed input site~$i_0$. 
Thus, the dynamics is fully deterministic and it is clear
that in the configuration space (recurrent configurations)
a DBTW model will settle down in an orbit with some period $T$, which 
in general will depend on $i_0$. 
For example, the period of a DBTW model with
a seeding site at the corner of the lattice is larger
than the period of the center-seeded DBTW model, as it is intuitively
clear and also known from computer simulations.  
An analytical approach~\cite{MARKOSOVA_1} was used to evaluate~$T$ 
for the center-seeded DBTW model
up to a system of $N=361$ sites, where the authors found a period of 
length $\approx 10^{17}$ and extrapolated their results
to reproduce the numerical estimation 
of~\cite{WIESENFELD_1} $T\sim \exp{(0.11\,N)}$.

In~\cite{WIESENFELD_1}, several interesting 
features of the DBTW model could be
derived by using the Abelian sandpile formalism:
\begin{itemize}
\item[(i)] The orbits have the same period (for a fixed input site $i_0$), 
           regardless of the initial conditions.
\item[(ii)] The minimal stable configuration 
           $C^{*}=\{z_i=z_c-1, {\rm \; for \; all}\;i\}$ is
           always on an orbit of the DBTW model. 
\end{itemize}
Note that (ii) does not mean that a DBTW model will always 
reach $C^{*}$ at some point. For example, for the 
initial condition $z_i=1$ for all~$i$ 
a system of $49$ sites has an orbit without $C^{*}$.
Only for different initial conditions or a different
system size (e.g.~$25$ sites with the same initial
condition) $C^{*}$ is on the orbit.

Let us now define $N_{{i_0}j}$ as the total number of
topplings at site $j$ within the period $T$ of a DBTW model 
with input site~$i_0$. 
The time average of $N_{{i_0}j}$ is simply 
$N_{{i_0}j}/T$. The total flow of particles during $T$ into $j$
has to equal the flow out of $j$ and it follows 
$N_{{i_0}j}/T={\Delta}^{-1}_{{i_0}j}$ or
\begin{equation}
{\Delta}^{-1}_{{i_0}j} = G_{i_0j} = N_{{i_0}j}/T.
\end{equation}
This means that every ''row'' $i$ of the BTW correlation function
$G_{ij}$, which stands for the ensemble-averaged number
of topplings in $j$ when seeding in $i$, represents
the time average of topplings in $j$ of a DBTW model 
with a fixed input site $i_0$.
The average avalanche size $\langle s \rangle$ for the 
BTW model can therefore be thought of as $\langle s_{i_0} \rangle$ 
of the DBTW models averaged over all orbits and
all possible seeding sites $i_0=1,2,...,N$.

Similar to the BTW model it is possible 
to calculate the average number of topplings~$\langle s \rangle$ 
for the center-seeded DBTW model 
on a square lattice with length~$L$ ($L$ odd). 
Using Eq.~(\ref{eq:corr}) we get with
$x_{i_0}= y_{i_0}= (L+1)/2$
\begin{eqnarray}
&\langle s \rangle &
 =  \sum \limits_{j} G_{i_0 j} \nonumber \\
& = & \frac{1}{{(L+1)}^{2}} \sum\limits^{\frac{L-1}{2}}_{a,b=0}
{(-1)}^{a+b} 
\frac{\cot{\frac{\tilde{A}}{2}}\; \cot{\frac{\tilde{B}}{2}}}
{\sin^2{\frac{\tilde{A}}{2}}+\sin^2{\frac{\tilde{B}}{2}}}
\label{eq:meansize_3}
\end{eqnarray}
with $\tilde{A}=(2a+1)\pi /(L+1)$
and $\tilde{B}=(2a+1)\pi /(L+1)$, respectively.
It is straightforward (though tedious) 
to evaluate this expression for large $L$ by using
standard methods.
One obtains $\langle s \rangle\sim L^2$, analogous to the BTW model. 
It is also easy to get an expression for the time-averaged number
of topplings $N_{{i_0}{i_0}}/T$ of the input site~$i_0$:
\begin{eqnarray}
\frac{N_{{i_0}{i_0}}}{T} 
&=& G_{{i_0}{i_0}}\nonumber\\
&=&\frac{1}{(L+1)^2}\sum\limits^{\frac{L-1}{2}}_{a,b=0}
\frac{1}{\sin^2{\frac{\tilde{A}}{2}}+\sin^2{\frac{\tilde{B}}{2}}}
\end{eqnarray}
which can be shown, again using elementary methods, to scale 
as $\frac{N_{{i_0}{i_0}}}{T}\sim\ln{L}$ for large $L$.

\section{Dynamics}
\label{sec:dynamics}

\subsection{Characterization of the Avalanches}
\label{subsec:avalanches}

\begin{figure}[t]
 \includegraphics[width=8cm]{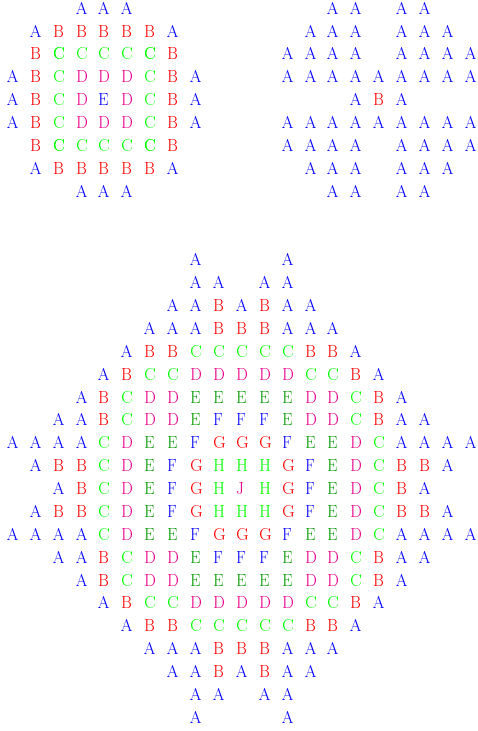}
 \caption{Three different avalanches for the center-seeded DBTW model.
          Sites which toppled once are marked as $A$, twice toppled
          sites as~$B$, etc. 
 \label{fig:cdbtw_avalanches}} 
\end{figure}

For a graphical representation of the avalanches, it is convenient 
to denote how many
topplings $n$ at each site have occurred during 
the avalanche~\cite{GRASS_1}. 
Sites with the same number
of topplings (during one avalanche) form ``shells''.
It is easy to check that for each site $i$ inside such 
a shell, for which all four neighbors are also
part of the shell, the height $z_i$ before and after the avalanche
remains the same.
Figure~\ref{fig:cdbtw_avalanches} shows some examples which have 
been obtained in the stationary state of the central seeded DBTW. 
The shells seem to form compact sets with $n$ monotonically
decreasing from the central site towards the boundaries along 
the symmetry axis.
Also, it seems that at the boundary of each $n={\rm const}$
shell, $n$ will change only by 1 (especially at the boundary of
the avalanche one finds $n=1$ most often). 
However, the latter statement is not always correct, there 
are rare exceptions. 
For instance, the upper left cluster in 
Fig.~\ref{fig:cdbtw_avalanches}
has a boundary site which has toppled  twice and
the lower cluster has sites with $n=1$ adjacent 
to a site with $n=3$. 
We could only prove two properties of the avalanches.
First, it has been shown for the BTW model, and the proof 
applies for the DBTW model also, that avalanches are always 
compact~\cite{GRASS_1,CHRIS_2}.
Second, one can show that $n$ decreases monotonically 
from the center towards the boundaries along the
symmetry axis by decomposing an avalanche into a series
of waves of topplings~\cite{IVASH_3,PRIEZ_2}. 
First one topples the center site and relaxes 
all other sites which become unstable. 
This defines the first wave of topplings. 
After this, one allows the center site to relax again
(if possible) which generates a second wave of topplings and
so forth. 
It has been shown~\cite{IVASH_3} that each avalanche 
produced by such a wave of topplings is compact. 
Furthermore, each site in a wave can topple only once. 
Thus, by superposition of the compact waves of topplings, 
$n$~can only monotonically decrease
from the center towards the boundaries along the 
symmetry axis.

\subsection{Scaling behavior of the avalanches}
\label{messung}

In driven systems the boundary conditions can 
influence the stationary state~\cite{KRUG}.
We show below that the scaling properties
of the BTW model and its deterministic
version are only the same if the latter has
boundary conditions or an initial configuration which
do not respect the square symmetry.

\begin{figure}[b]
 \includegraphics[width=8.6cm]{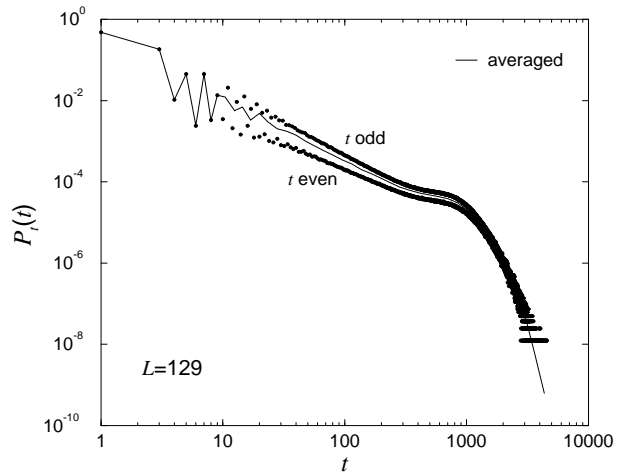}
 \caption{The probability distribution of 
          the avalanche duration~$P_t(t)$.
          Avalanches of an even or odd duration
          display a different scaling behavior.
          Thus, the usual logarithmic averaging of
          the distribution leads to useless results
          (solid line). 
 \label{fig:cdbtw_duration}} 
\end{figure}

We denote by~$s$ the total number of
topplings which occurred during the lifetime~$t$
(in units of lattice sweeps) of an avalanche. 
The area~$a$ of such an avalanche is the number of
distinct toppled sites.
The outflow~$o$ is the total number of sand grains 
which leave the system during an avalanche.
The linear size of an avalanche is measured via the radius
of gyration of an avalanche cluster.
In the critical steady state the corresponding probability
distributions should obey a power-law behavior 
\begin{equation}
P_x(x) \; \sim \;  x^{-\tau_x},
\label{eq:prob_dist}
\end{equation}
characterized by the exponent $\tau_x$ with
$x \in \{s,a,t,o,r\}$.
As usual, we measure the avalanche distributions
by counting the numbers of avalanches corresponding 
to a given area, duration, etc.~and 
integrate these numbers over bins of increasing
length (see for instance~\cite{MANNA_2}).
In our simulations successive bin lengths 
increase by a factor $b=1.2$.
In the case of the BTW model it is known
that the probability distributions display logarithmic corrections
\begin{equation}
P_x(x) \; \sim \;  x^{-\tau_x}\, x^{{\rm const}/\ln{L}},
\label{eq:prob_dist_log_corr}
\end{equation}
which are caused by finite-size effects~\cite{MANNA_1,LUEB_2}.
A numerical determination of the avalanche exponents
requires therefore a careful analysis of finite-size effects.
Using the functional form of the finite-size corrections
it is possible to determine the exponents~$\tau_x$
directly, i.e.~without any extrapolation to the
infinite system, and the best known values are 
$\tau_s = 1.293 \pm 0.009$, $\tau_a = 1.33 \pm 0.011$,
$\tau_t = 1.480 \pm 0.011$, 
and $\tau_r = 1.665 \pm 0.013$~\cite{LUEB_2}.

\begin{figure}[t]
 \includegraphics[width=8.6cm]{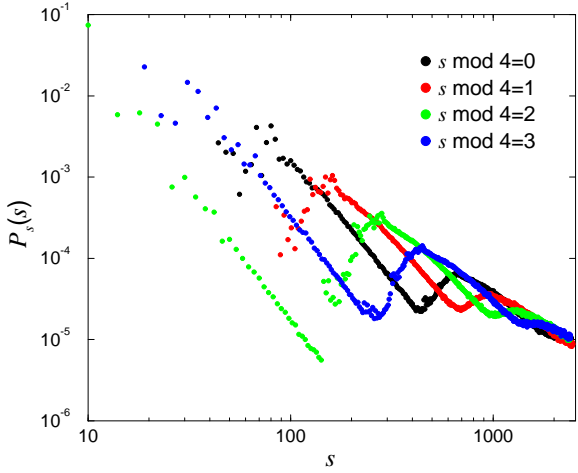}
 \caption{The probability distribution of 
          the avalanche duration~$P_s(s)$.
          The distribution decomposes into four
          different branches corresponding to the 
          four possible values of $s\, {\rm mod}\,4$.
 \label{fig:cdbtw_size}} 
\end{figure}

\begin{figure}[b]
 \includegraphics[width=8.6cm]{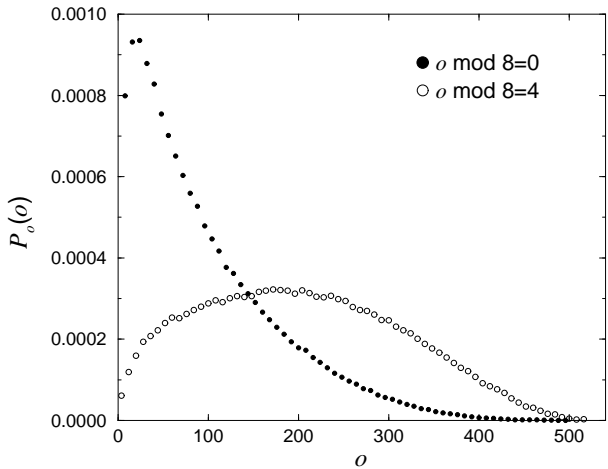}
 \caption{The probability distribution of 
          the sand outflow~$P_o(o)$ for $L=129$.
          The distribution decomposes into two
          different branches corresponding to the 
	  two possible values of $o\, {\rm mod}\,8$.
\label{fig:cdbtw_outflow}}
\end{figure}

We performed simulations of the center-seeded DBTW model 
for various system sizes $L\le 2049$ and averaged
all measurements over at least $2\times 10^6$ avalanches.
Starting from an empty lattice, we added 
particles at the lattice center and 
applied the ``burning algorithm'' 
in order to check if the system has
reached the steady state~\cite{DHAR_2}.
Thus the initial conditions (as well as the
boundary conditions) respect the square symmetry.
Figure~\ref{fig:cdbtw_duration} shows the probability distribution
$P_t(t)$ of the avalanche duration.
Surprisingly, it turns out that the above
mentioned logarithmically averaging method
is not suitable in our case, because
avalanches of odd or even duration display a 
different scaling behavior. 
The two branches of $P_t(t)$ 
in Fig.~\ref{fig:cdbtw_duration} ($\tau_t^{\rm odd}$ and
$\tau_t^{\rm even}$)
clearly have different slopes for the system size
considered.
The probability distribution of the avalanche
size~$P_s(s)$ exhibits an even more
complicated fine structure
of four distinct branches corresponding to
the four possible values of $s\,{\rm mod}\,4$~(see
Fig.~\ref{fig:cdbtw_size}).
We have no explanation for this behavior.

\begin{figure}[b]
\includegraphics[width=8.6cm]{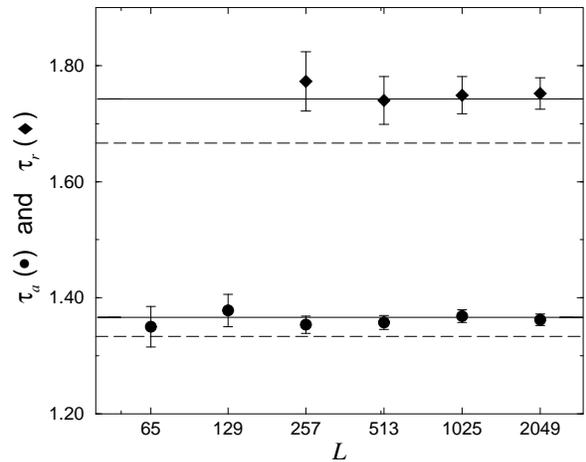}
 \caption{System size dependence of the avalanche
          exponents $\tau_a$ and $\tau_r$.
          The solid lines corresponds to the 
          values obtained from a finite-size scaling
          analysis and the dashed lines corresponds
          to the values of the BTW model obtained 
          in~\protect\cite{LUEB_2}.
 \label{fig:cdbtw_tau_reg}} 
\end{figure}

Let us briefly remark that a similar symmetry effect can be
found in the outflow probability distribution $P_o(o)$.
One gets different curves for all $o$ which are
divisible by 8 and those which are not (see
Fig.~\ref{fig:cdbtw_outflow}).
To explain this, we note that the outflow~$o$ is divisible
by 8 exactly when the sites in the middle of the 
boundary edges do {\it not} topple. 
This is extremely unlikely for large avalanches. 
On the other hand, for small
avalanches, the constraint to topple sites in the middle
of the boundary edges considerably reduces the number of
possible avalanches.

\begin{figure}[t]
 \includegraphics[width=8.6cm]{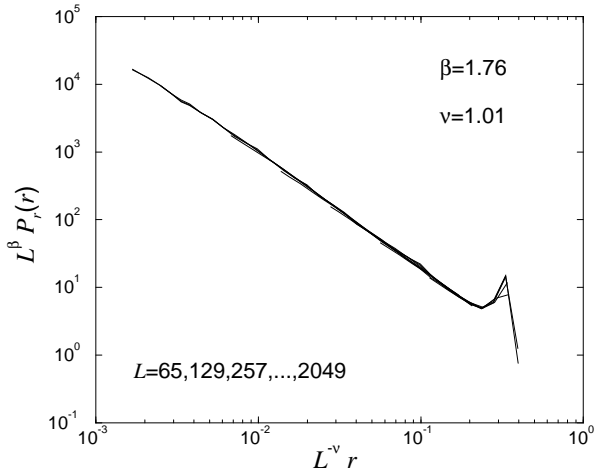}
 \caption{The finite-size scaling analysis
          of the avalanche distribution~$P_r(r)$.
          Since the radius scales with the
          system size the exponent $\nu_r$ 
          should equal one.
\label{fig:cdbtw_fss_radius}}
\end{figure}

Since the exponents of the size and 
duration distribution, $\tau_s$ and $\tau_t$,
are not well defined here we renounce further 
investigations of the size and duration
distribution in this section and focus our
attention on the probability distributions of the
avalanche area and radius which behave as usual (see below).
We measured the probability distributions
$P_a(a)$ and $P_r(r)$ for various system
sizes and obtained the corresponding exponents
from a power-law fit of the straight portion
of the curves. 
The values of both exponents are plotted in
Fig.~\ref{fig:cdbtw_tau_reg}.
In contrast to the BTW model [Eq.(\ref{eq:prob_dist_log_corr})] 
the avalanche exponents
of the center-seeded DBTW model display
no significant system size dependence.
This allows us to apply the finite-size
scaling analysis~\cite{KADANOFF}
\begin{equation}
P_x(x,L) \; = \;
L^{-\beta_x} \, g_x ( x L^{-\nu_x} ), 
\label{eq:fss_simple}
\end{equation}
where the scaling exponents~$\beta_x$ and~$\nu_x$ 
are connected with the avalanche exponent
$\tau_x$ via the scaling equation
$\beta_x = \tau_x \nu_x$~\cite{KADANOFF}.
This finite-size scaling ansatz works for
the area and radius distribution and the corresponding data
collapse for $P_r(r)$ is plotted 
in Fig.~\ref{fig:cdbtw_fss_radius}.
The obtained values for the avalanche exponents agree 
with the results of the regression analysis 
(see Fig.~\ref{fig:cdbtw_tau_reg})
and we get $\tau_a=1.368\pm0.011$ and
$\tau_r=1.752\pm0.027$.

As already mentioned in section~\ref{subsec:avalanches}, 
it is possible to show that the avalanches
are compact.
Thus, the area scales with the radius as
\begin{equation}
a \; \sim \; r^2.
\label{eq:area_radius}
\end{equation}
Then, the transformation law of probability distributions 
$P_a(a) \mbox{d}a=P_r(r) \mbox{d}r$ leads to
the scaling relation
\begin{equation}
{2}\;=\;\frac{\tau_r-1}{\tau_a-1}.
\label{eq:gam_ar}
\end{equation}
This scaling relation is fulfilled 
within the error-bars, which confirms
the accuracy of the determination of the
avalanche exponents $\tau_a$ and $\tau_r$.
Finally we mention that our obtained
exponents are consistent with the
values $\tau_a=11/8$ and $\tau_t=7/4$
and that these values obey the above scaling relation
exactly.

\section{Discussion}

Since both avalanche exponents differ significantly 
from those of the BTW model 
we conclude
that the center-seeded DBTW model does not
belong to the BTW universality class.
It is worth to examine the different 
universal behavior in detail
since the universality hypothesis
of Ben-Hur and Biham states that the only 
parameter which determines the scaling
behavior (exponents) of a sandpile model
is the so-called relaxation vector which
describes how the sand grains of a critical
site are distributed to the next 
neighbors~\cite{BENHUR}.
Applying this concept of classification 
one can identify three universality classes 
where the distribution is {\it nondirected},
{\it nondirected on average}, and {\it directed}.
For instance, the BTW and the related Zhang
model~\cite{ZHANG_1} belong to the 
universality class of nondirected models,
whereas the Manna model~\cite{MANNA_2} is nondirected on
average and therefore belongs to a different class.
Several numerical investigations confirm
these classification ansatz~(see for 
instance~\cite{BENHUR,LUEB_2,LUEB_3} and 
for recent investigations~\cite{CHESSA_2,LUEB_9}).

\begin{figure}[t]
 \includegraphics[width=8.6cm]{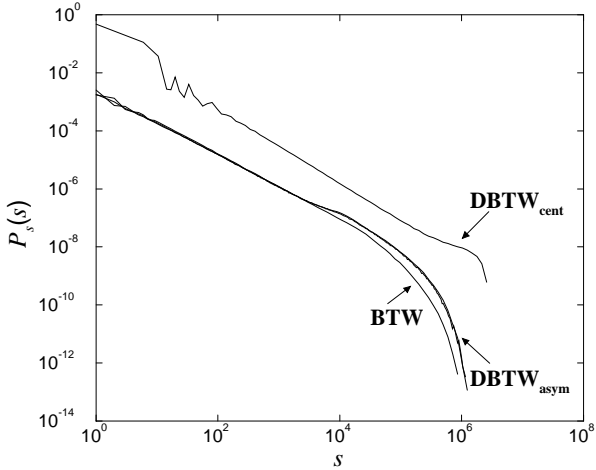}
 \caption{The probability distribution $P_s(s)$ of the
          symmetric center-seeded DBTW model, two asymmetric
          DBTW models (in one case the square symmetry is
          broken by the boundaries and in the other case
	  by asymmetric initial conditions),
	  and the BTW model for $L=257$.
          In the latter cases the curves
          are shifted in the downward direction.
\label{fig:adbtw_size_dist_01}}
\end{figure}

\begin{figure}[t]
 \includegraphics[width=8.6cm]{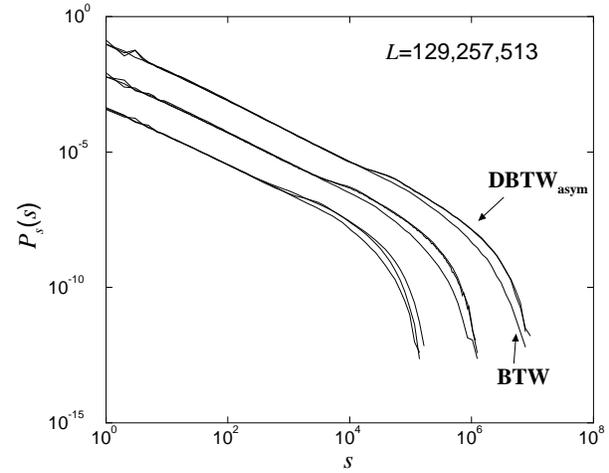}
 \caption{The probability distribution $P_s(s)$ of 
          the two asymmetric DBTW models and the 
          BTW model for various system sizes.
          For $L< 513$ the curves are shifted in the
          downward direction.
\label{fig:adbtw_size_dist_02}}
\end{figure}

According to this classification concept
the center-seed\-ed DBTW model
and the BTW model should belong to the same universality
class because both models are characterized by the
same relaxation vector.
In contrast to the BTW model the center-seeded DBTW model 
is deterministic and both the avalanches and the 
height configurations display the square symmetry.
Moving the input site~$i_0$ from the lattice center 
brakes the square symmetry but the dynamics of the
system is still deterministic. 
The same effect is obtained if we use center-seeding 
but start with an asymmetric initial condition.
We call these models the asymmetric DBTW models
and our analysis revealed that they display
the same scaling behavior as the usual BTW model.
We plot the probability distribution $P_s(s)$ for the 
considered models in Fig.~\ref{fig:adbtw_size_dist_01}.
Except of deviations at the cut-off, the probability
distributions of the asymmetric DBTW models agree with
the corresponding curve of the BTW model
and differ clearly from the distribution
of the symmetric DBTW model.
Figure~\ref{fig:adbtw_size_dist_02} shows the
probability distribution for various system
sizes.
Again, apart from the cut-off behavior the curves
of the BTW model and the asymmetric DBTW models
are identical.
This implies that the avalanche exponents of the 
asymmetric DBTW models display the same logarithmic
corrections as the BTW model~[Eq.~(\ref{eq:prob_dist_log_corr})].

We conclude from our investigations
that the asymmetric DBTW and the BTW model belong
to the same universality class,
whereas the center-seeded DBTW model with symmetric
initial height configuration does not.
Since the asymmetric DBTW model is still deterministic
but lacks the square symmetry 
the different universal behavior of the 
center-seeded DBTW model is not caused
by the deterministic dynamics but by the 
square symmetry of the system (in agreement 
with~\cite{WIESENFELD_1}).

We conclude from our results
that properties of the steady state such as
symmetries or translational invariance 
can affect the universality class of sandpile models.
This is confirmed by recently performed simulations~\cite{LUEB_UNPUB}
of a directed version of the Zhang model
which exhibits a different scaling behavior
than the exactly solved directed BTW model~\cite{DHAR_1}.
According to the classification of Ben-Hur
and Biham the directed Zhang model and the
directed BTW model should
belong to the same universality class.
But in contrast to the directed BTW model
the height configuration of the directed
Zhang model displays no translation invariance~\cite{LUEB_UNPUB}.
Similar to the center-seeded DBTW model 
one has to be careful to apply the universality
hypothesis of Ben-Hur and Biham. 

In summary we reconsidered a deterministic version of the
Bak-Tang-Wiesenfeld sandpile model where the sand grains
are added always to the central site of the lattice.
Similar to the usual BTW model the 
{\it Abelian sandpile} formalism allows
to calculate some of the static 
properties of the system.
Our numerical investigations show that the 
deterministic central-seeded model 
with square symmetric initial conditions exhibits
a different scaling behavior than the 
BTW model, in contrast to the deterministic model
without square symmetry.

\section{Acknowledgments}
NR gratefully acknowledges a postdoctoral fellowship from the Deutsche
Forschungsgemeinschaft and thanks Joel Le\-bo\-witz for hospitality at the
Mathematics Department of Rutgers University and for support under NSF
grant DMR~95--23266 and DIMACS.

\end{document}